\documentclass[onecolumn,showpacs,preprintnumbers,amsmath,amssymb,prd, superscriptaddress]{revtex4-1}

\usepackage{graphicx}
\usepackage{dcolumn}
\usepackage{bm}

\begin{document}


\title{Constraints on the Extremely-high Energy Cosmic
Neutrino Flux \\with the IceCube 2008-2009 Data}
\affiliation{III. Physikalisches Institut, RWTH Aachen University, D-52056 Aachen, Germany}
\affiliation{Dept.~of Physics and Astronomy, University of Alabama, Tuscaloosa, AL 35487, USA}
\affiliation{Dept.~of Physics and Astronomy, University of Alaska Anchorage, 3211 Providence Dr., Anchorage, AK 99508, USA}
\affiliation{CTSPS, Clark-Atlanta University, Atlanta, GA 30314, USA}
\affiliation{School of Physics and Center for Relativistic Astrophysics, Georgia Institute of Technology, Atlanta, GA 30332, USA}
\affiliation{Dept.~of Physics, Southern University, Baton Rouge, LA 70813, USA}
\affiliation{Dept.~of Physics, University of California, Berkeley, CA 94720, USA}
\affiliation{Lawrence Berkeley National Laboratory, Berkeley, CA 94720, USA}
\affiliation{Institut f\"ur Physik, Humboldt-Universit\"at zu Berlin, D-12489 Berlin, Germany}
\affiliation{Fakult\"at f\"ur Physik \& Astronomie, Ruhr-Universit\"at Bochum, D-44780 Bochum, Germany}
\affiliation{Physikalisches Institut, Universit\"at Bonn, Nussallee 12, D-53115 Bonn, Germany}
\affiliation{Dept.~of Physics, University of the West Indies, Cave Hill Campus, Bridgetown BB11000, Barbados}
\affiliation{Universit\'e Libre de Bruxelles, Science Faculty CP230, B-1050 Brussels, Belgium}
\affiliation{Vrije Universiteit Brussel, Dienst ELEM, B-1050 Brussels, Belgium}
\affiliation{Dept.~of Physics, Chiba University, Chiba 263-8522, Japan}
\affiliation{Dept.~of Physics and Astronomy, University of Canterbury, Private Bag 4800, Christchurch, New Zealand}
\affiliation{Dept.~of Physics, University of Maryland, College Park, MD 20742, USA}
\affiliation{Dept.~of Physics and Center for Cosmology and Astro-Particle Physics, Ohio State University, Columbus, OH 43210, USA}
\affiliation{Dept.~of Astronomy, Ohio State University, Columbus, OH 43210, USA}
\affiliation{Dept.~of Physics, TU Dortmund University, D-44221 Dortmund, Germany}
\affiliation{Dept.~of Physics, University of Alberta, Edmonton, Alberta, Canada T6G 2G7}
\affiliation{Dept.~of Physics and Astronomy, University of Gent, B-9000 Gent, Belgium}
\affiliation{Max-Planck-Institut f\"ur Kernphysik, D-69177 Heidelberg, Germany}
\affiliation{Dept.~of Physics and Astronomy, University of California, Irvine, CA 92697, USA}
\affiliation{Laboratory for High Energy Physics, \'Ecole Polytechnique F\'ed\'erale, CH-1015 Lausanne, Switzerland}
\affiliation{Dept.~of Physics and Astronomy, University of Kansas, Lawrence, KS 66045, USA}
\affiliation{Dept.~of Astronomy, University of Wisconsin, Madison, WI 53706, USA}
\affiliation{Dept.~of Physics, University of Wisconsin, Madison, WI 53706, USA}
\affiliation{Institute of Physics, University of Mainz, Staudinger Weg 7, D-55099 Mainz, Germany}
\affiliation{Universit\'e de Mons, 7000 Mons, Belgium}
\affiliation{Bartol Research Institute and Department of Physics and Astronomy, University of Delaware, Newark, DE 19716, USA}
\affiliation{Dept.~of Physics, University of Oxford, 1 Keble Road, Oxford OX1 3NP, UK}
\affiliation{Dept.~of Physics, University of Wisconsin, River Falls, WI 54022, USA}
\affiliation{Oskar Klein Centre and Dept.~of Physics, Stockholm University, SE-10691 Stockholm, Sweden}
\affiliation{Dept.~of Astronomy and Astrophysics, Pennsylvania State University, University Park, PA 16802, USA}
\affiliation{Dept.~of Physics, Pennsylvania State University, University Park, PA 16802, USA}
\affiliation{Dept.~of Physics and Astronomy, Uppsala University, Box 516, S-75120 Uppsala, Sweden}
\affiliation{Dept.~of Physics, University of Wuppertal, D-42119 Wuppertal, Germany}
\affiliation{DESY, D-15735 Zeuthen, Germany}

\author{R.~Abbasi}
\affiliation{Dept.~of Physics, University of Wisconsin, Madison, WI 53706, USA}
\author{Y.~Abdou}
\affiliation{Dept.~of Physics and Astronomy, University of Gent, B-9000 Gent, Belgium}
\author{T.~Abu-Zayyad}
\affiliation{Dept.~of Physics, University of Wisconsin, River Falls, WI 54022, USA}
\author{J.~Adams}
\affiliation{Dept.~of Physics and Astronomy, University of Canterbury, Private Bag 4800, Christchurch, New Zealand}
\author{J.~A.~Aguilar}
\affiliation{Dept.~of Physics, University of Wisconsin, Madison, WI 53706, USA}
\author{M.~Ahlers}
\affiliation{Dept.~of Physics, University of Oxford, 1 Keble Road, Oxford OX1 3NP, UK}
\author{K.~Andeen}
\affiliation{Dept.~of Physics, University of Wisconsin, Madison, WI 53706, USA}
\author{J.~Auffenberg}
\affiliation{Dept.~of Physics, University of Wuppertal, D-42119 Wuppertal, Germany}
\author{X.~Bai}
\affiliation{Bartol Research Institute and Department of Physics and Astronomy, University of Delaware, Newark, DE 19716, USA}
\author{M.~Baker}
\affiliation{Dept.~of Physics, University of Wisconsin, Madison, WI 53706, USA}
% ...

\date{\today}
\collaboration{IceCube Collaboration}\noaffiliation

\begin{abstract}
 We report on a search for extremely-high energy neutrinos with
 energies greater than $10^6$~GeV using the data taken with the IceCube
 detector at the South Pole. The data was collected between April 2008 and
 May 2009 with the half completed IceCube array.
 The absence of signal candidate events in the sample of 333.5 days of 
 livetime significantly improves model independent limit from previous searches
and allows to place a limit on the diffuse flux of cosmic neutrinos with an $E^{-2}$ spectrum 
in the energy range $2.0 \times 10^{6}$ $-$ $6.3 \times 10^{9}$~GeV to a level of $E^2 \phi \leq 3.6 \times 10^{-8}$~${\rm GeV cm^{-2} sec^{-1}sr^{-1}}$.
\end{abstract}

{\large
This version of astro-ph:1103.4250 contains the original (published) version of this article
(Phys. Rev. D 83, 092003 (2011)), as well as its erratum. The original document has not
been modified, but the reader should use the effective area values from the erratum.
}
\newpage
\maketitle
\section{\label{sec:intro} Introduction}
Cosmogenic neutrinos, the daughter particles of the 
Greisen-Zatsepin-Kuzmin (GZK) process in which the highest energy
cosmic-rays interacting with the cosmic-microwave background
\cite{GZK,berezinsky69}, may give a unique picture of the Universe 
in the highest energy regime. Cosmogenic neutrinos carry information 
about the sources of the highest energy cosmic-rays, 
such as their location, cosmological evolution, and cosmic-ray spectra 
at the sources.    
Various cosmogenic neutrino models~\cite{yoshida93,kalashev02,ESS,
  ahlers} which assume primary cosmic-ray protons predict neutrino fluxes $E_{\nu}^2 \phi \geq 10^{-4}$~${\rm
  GeV m^{-2} sec^{-1} sr^{-1}}$ in the energy range $10^8$~GeV $\leq
E_{\nu} \leq 10^{10}$~GeV, which implies that the $4\pi$ solid angle
averaged neutrino effective area divided by energy
${A_{\nu}}/{E_{\nu}}$ must be larger than $10^{-5}$~m$^2$/GeV
(e.g.\ $A_{\nu} \geq 10^3$m$^2$ at $10^8$~GeV and $A_{\nu} \geq
10^4$m$^2$ at $10^9$~GeV) to detect several cosmogenic neutrinos every
year.

Several techniques have been used to realize such huge detection volumes for
these extremely-high energy (EHE) neutrinos.
Air-shower detectors search for neutrino induced young inclined  
showers~\cite{hires} or Earth-skimming events initiated by tau neutrinos~\cite{auger}.
Radio Cherenkov neutrino detectors search for radio Askar'yan pulses in a
dielectric medium as the EHE neutrino signature~\cite{anita, rice,
  glue}. Underground neutrino telescopes, such as IceCube, deployed in transparent
naturally occurring media~\cite{amanda, ic22} can detect EHE neutrino
interactions through the strong Cherenkov radiation emitted by the
charged secondary particles. This technique is well established for
observations of astrophysical neutrinos in the MeV to GeV energy
region~\cite{sk,sno}, and can also be utilized to search for cosmogenic
EHE neutrinos with an appropriate background rejection method. In a
neutrino telescope, an EHE neutrino interaction would be identified by
the extremely high number of Cherenkov photons deposited in the
detector. 

In this paper, we describe the search for neutrinos with energies
above $10^6$~GeV using data collected with the half-completed IceCube
detector in 2008$-$2009. This analysis is sensitive to all three
neutrino flavors. Compared to the previous EHE neutrino search
described in Ref.~\cite{ic22}, which used an earlier stage of 
the IceCube detector, the current analysis benefits from  
the enlarged instrumented volume
and from improved agreement between simulated and observed event
distributions. 
This article presents the improved strategies implemented since the previous analysis~\cite{ic22}.
\section{\label{sec:dataset} Data Sets}

The analysis uses data collected from April 6, 2008 through May
20, 2009.
At the time of data collection, the IceCube detector consisted of 2400
Digital Optical Modules (DOMs) on 40 vertical strings. 
The volume of the detector was roughly 0.5~km$^3$ with
the detector center located at a depth of 1948~m below the ice surface.
The DOMs consist of a 25~cm photomultiplier tube (PMT) \cite{pmt} with
data acquisition and calibration electronics, data compression,
communications, and control hardware~\cite{digitizer}. The trigger
setting was unchanged from the previous analysis~\cite{ic22}.

The analysis was optimized on simulated data with most of the experimental
data kept blind. A 10\% subset of the experimental data was used
for examinations of the Monte Carlo simulations and detector
response. This subset comprised 35.8 days of detector livetime
distributed randomly throughout the data collection period, and was
not used once the analysis was fully defined.
The use of statistically independent final sample conservatively ensures avoidance of possible analysis bias due to tuning a Monte Carlo simulation using an experimental subset.
 The selection criteria were then applied to the complementary 90\% of the
experimental data, comprising 333.5 days of livetime.

The primary background in this analysis is muon bundles made up of 
large numbers of muons produced by high energy cosmic-ray interactions
in the atmosphere. This background was simulated with the {\sc
  corsika} air-shower simulation package version 6.720~\cite{corsika} 
with the {\sc sibyll 2.1}~\cite{sibyll} and {\sc qgsjet-ii}~\cite{qgsjetii} 
hadronic interaction models, without prompt muons from the heavy meson
decays. Cosmic-ray interactions assuming pure proton and iron primary
compositions in the energy region between $10^6$ and $10^{10}$~GeV
were simulated. Background contributions from primary cosmic-ray
energies beyond $10^{10}$~GeV were estimated by extrapolation of the
simulated sample up to the GZK cutoff energy of $\sim$5 $\times
10^{10}$~GeV\null. EHE neutrino signal events in energies between $10^5$
and $10^{11}$~GeV from several flux models~\cite{yoshida93,kalashev02,ESS, ahlers, yoshida98, sigl} were
simulated using the {\sc juli}e{\sc t} package~\cite{yoshida04}.

\section{\label{sec:selection} Event Selection}
The amount of energy deposited in the form of Cherenkov photons by the
neutrino-induced charged particles in the detector is highly
correlated with the energy of the particles~\cite{ic22}. 
An EHE neutrino interaction occurring inside 
or close to the IceCube detector would stand out against the background of 
cosmic-ray induced muons due to the much higher light deposition.
The total number of photo-electrons (NPE)
recorded in an event was used as the main distinctive feature to
separate signal from background.

\subsection{On-line sample}
The number of photo-electrons (p.e.) recorded by an individual DOM was
derived by integrating the pedestal subtracted waveforms. Each DOM has
two waveform digitizers, that simultaneously
capture p.e.\ signals with differing dynamic ranges and time
windows~\cite{digitizer}. The event total NPE was then obtained by
summing the number of p.e.\ detected by each DOM.
PMT saturation effects and the sizes of the time windows limit the NPE estimation at high light levels.
The initial NPE calculation was performed online at the South
Pole. For this analysis we consider only events with
${\rm{NPE_{online}}}$~$\geq$~630. The event rate of this ``on-line
  bright sample'' was $\sim$1.4~Hz. At this level, the background rate
exceeded the expected signal rate by $\geq O(10^7)$.

\subsection{Off-line sample}
For the following data selection step, the NPE values were re-calculated after
eliminating photon signals from low energy muons accidentally
coincident in a 20~$\mu$s time window of a large NPE event. 
These low
energy muons leave a faint light, typically with an NPE~$<$~9. The light
deposition of the coinciding low energy muon was, in most cases,
spatially and temporally separated from the main bright p.e.\ cluster. 
%
While the few coincident photons have very small impact on the NPE calculation, they can disturb the geometrical reconstruction of the particle tracks later on in the analysis.
Contributions from coincident 
low energy muons were eliminated by removing p.e.\ signals that were 
temporally separated from the time of the highest light deposition 
associated with the main high NPE event. The recording time of a p.e.\ 
signal in the $i_{th}$ DOM, $t_{10,i}$, was defined as the time at which 10\% of the 
total charge had been captured. The time of the highest light deposition 
was defined as the time ${\rm t_{LN}}$ of the DOM which captured the largest p.e.\ 
signal in the event. This time, ${\rm t_{LN}}$, was typically associated with the 
time of closest approach of the charged particle tracks to any DOM in the 
detector. For the off-line NPE calculation and track reconstruction, those 
p.e.\ signals which occurred outside the time window [$-$4.4~$\mu$s, 6.4~$\mu$s] 
around the ${\rm t_{LN}}$ were excluded.
%
%
\begin{figure*}
  \includegraphics[height=1.6in, width=1.6in]{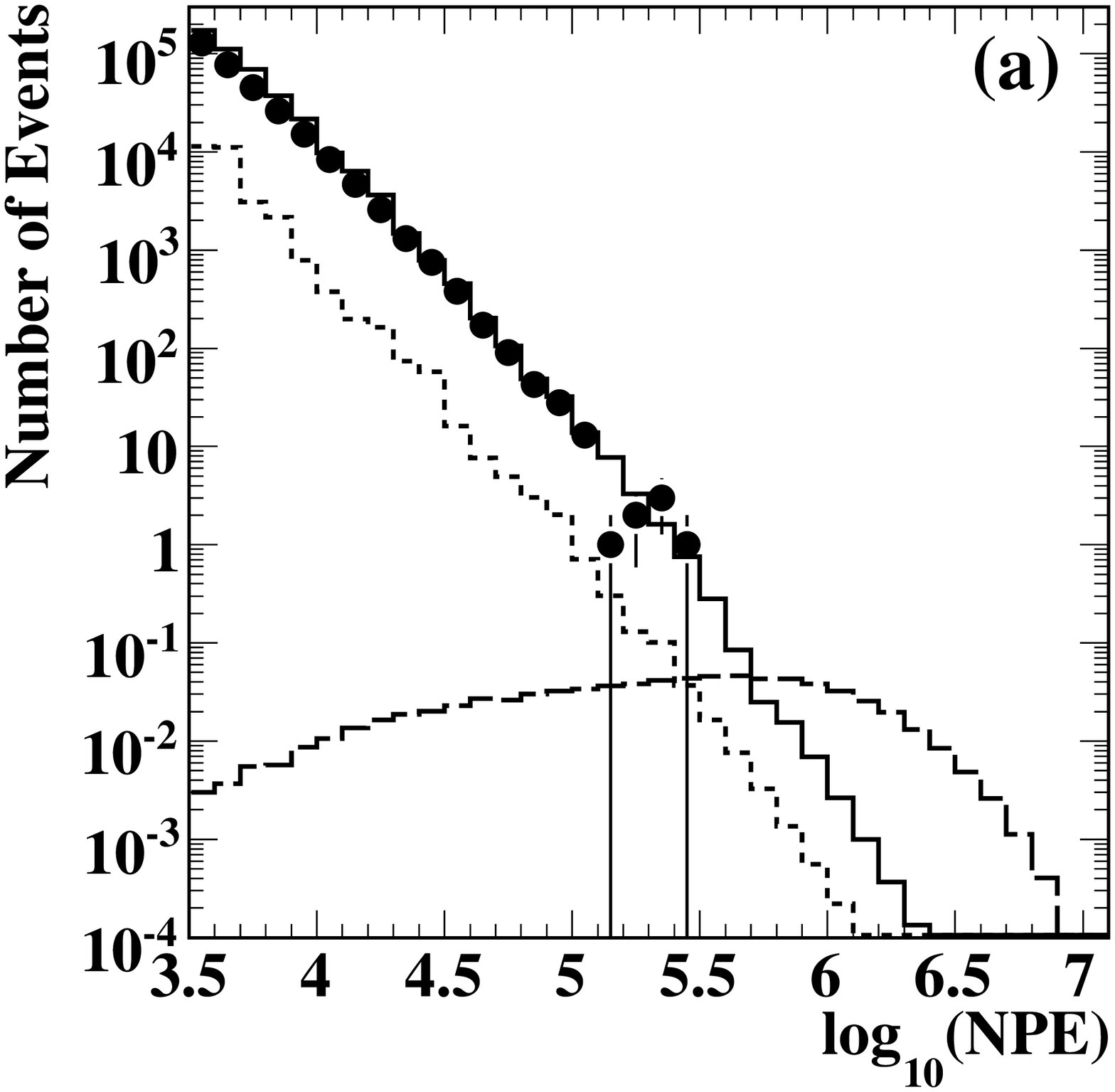}
  \includegraphics[height=1.6in, width=1.6in]{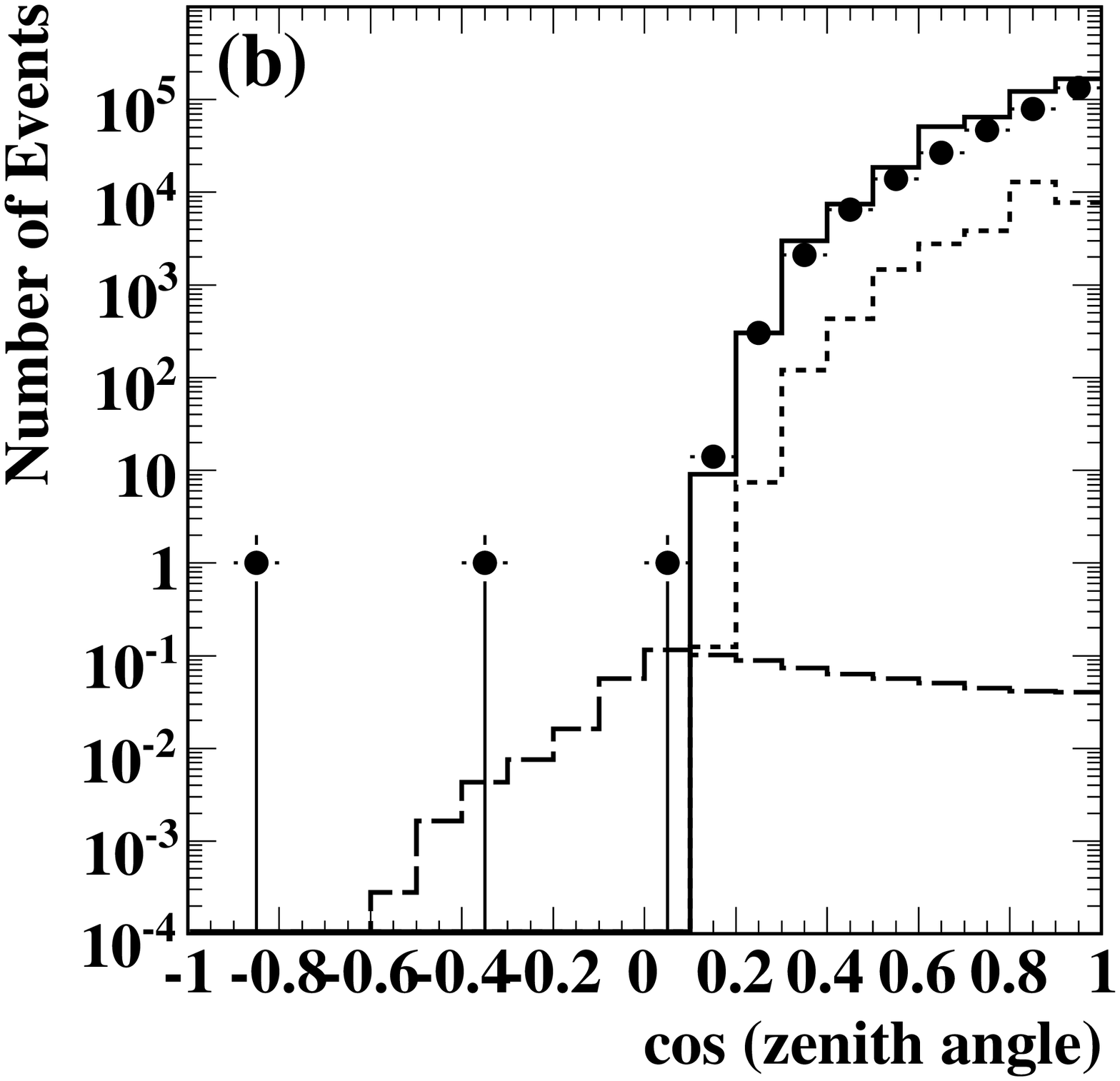}
  \includegraphics[height=1.6in, width=1.6in]{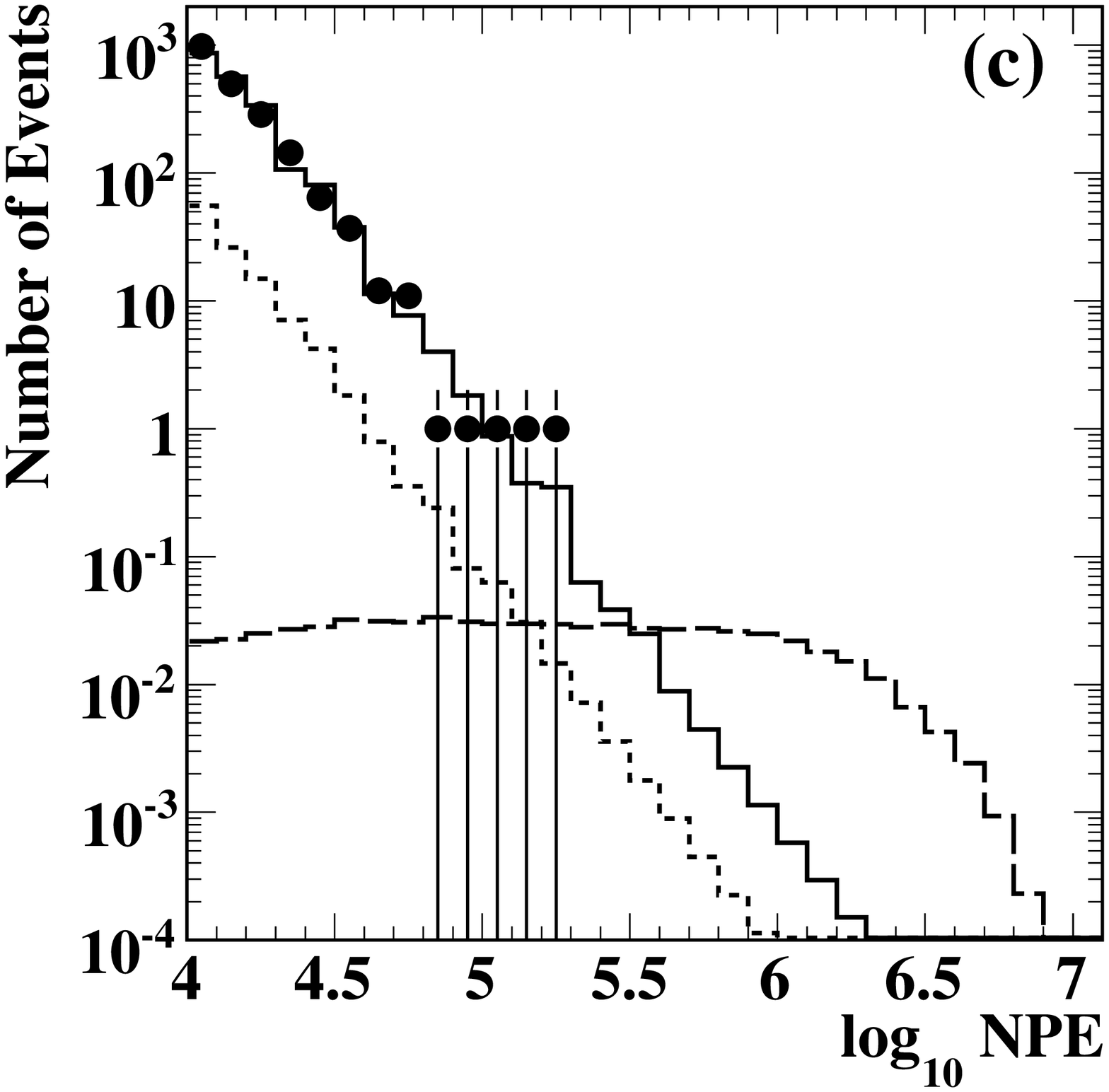}
  \includegraphics[height=1.6in, width=1.6in]{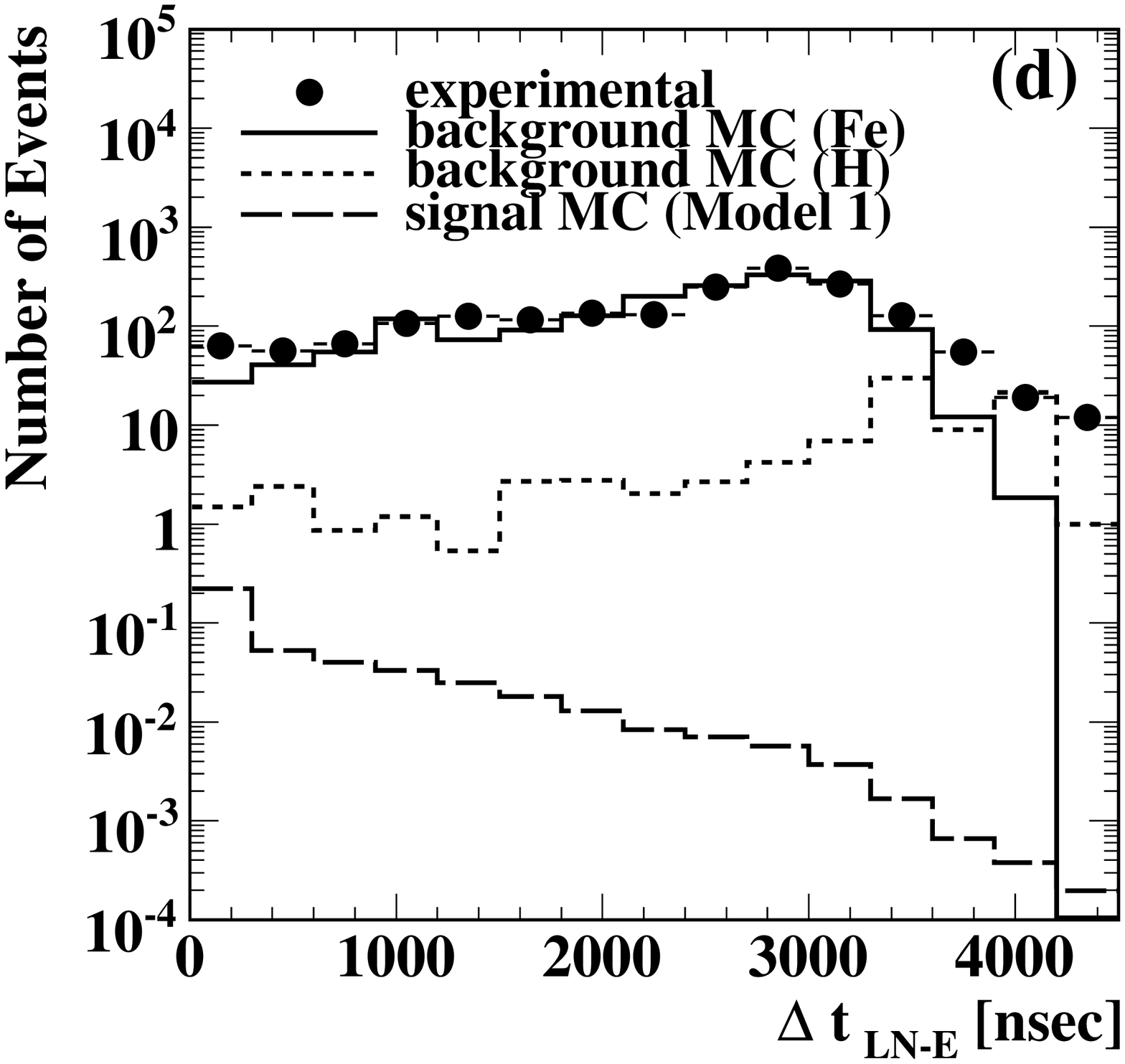}
  \caption{Event observables in the quality bright 
sample that are used for the final selection criteria.
Distributions of (a) NPE and (b) cosine of the 
reconstructed zenith angle for shallow events, and (c) NPE and (d) ${\rm \Delta t_{LN-E}}$ for deep events in a livetime of 333.5~days.
    The black circles represent experimental data and the solid and 
    dashed lines are {\sc corsika-sibyll} with iron and proton primaries, 
    respectively. 
    The expected signal distributions from 
simulations of the GZK~1 model (sum of all three neutrino flavors) are 
shown as long-dashed histograms. 
Systematic uncertainties are not included.
    \label{fig:level2}}
\end{figure*}
The ``off-line bright sample'' selects events with NPE $\geq 3.2
\times 10^3$ and the number of hit DOMs (NDOM)~$\ge 200$; here and
below NPE and NDOM are obtained after the ${\rm t}_{LN}$ time window
cleaning. These NPE and NDOM thresholds reduced the background rate by
two orders of magnitude while keeping $\sim$70~\% of the cosmogenic neutrino-induced events remain. 
The remaining backgrounds are bundles containing many hundreds of muons, with an estimated cosmic-ray energy above $10^7$~GeV\null.

\subsection{Quality cut}

Apart from NPE, the particle direction and the depth distribution of the 
detected Cherenkov photons are distinctive event features that separate the 
EHE neutrino signal from the atmospheric muon background.
Due to the energy dependence of the neutrino interaction cross section, 
most of the EHE neutrino signal is expected from directions close to the 
horizon. As a result of the depth dependence of the optical properties of 
the Polar ice, the largest photon signals are often detected in the 
deepest part of the detector, where the ice is most transparent~\cite{ice}.
On the other hand, the background atmospheric muons enter the detection 
volume from above and lose a substantial fraction of their energy during 
propagation through the detector. Therefore, the time and 
depth coordinates, z, of the detected Cherenkov photons, measured relative to the detector center, show negative correlation for background.
The largest photon signals from these background muons are expected 
at shallow depths near the top of the detector.
Exceptions are inclined atmospheric muon bundles that pass outside the 
instrumented volume with the point of closest approach in the deep, clear 
ice at the bottom of the detector, or individual muons that deposit most 
of their energy in an isolated catastrophic energy loss in the deep ice 
after having passed through the top part of the detector.
Track reconstructions often fail to identify such atmospheric muon events 
as downward-going tracks, when most of the light deposition occurs in the 
deep part of the detector. Therefore, a track reconstruction is applied 
only to those events in which the DOM with the largest signal is located 
at z $>$ $-$300~m (``shallow events"). The negative z value indicates the vertical distance below the center of the IceCube detector. 
For events with the largest 
photon signal at z $<$ $-$300~m (``deep events"), further event selection 
criteria rely on timing instead of directional information.

For the shallow events, the particle directions are reconstructed with
the LineFit algorithm~\cite{ic22}. Since the majority of the EHE
neutrino induced events is close to the horizon~\cite{yoshida04} while
the directions of the background muon bundles are mostly vertical, it is
important to minimize the number of background tracks that are
misreconstructed as horizontal. 
In order to reject the mis-reconstructed background events, another simple one-dimensional reconstruction is
introduced.
The distribution of average depth of p.e.\ as a function of timing are
fitted by a linear function, $\overline{z}(t_{10}) = C_0 + {\rm
  S_{zt}} \cdot t_{10}$. 
%
The fit parameter, ${\rm S_{zt}}$, 
is a measure for the speed at which the light signal propagates in 
z-direction, and hence for the inclination of the tracks. For vertically 
downward-going relativistic particles, the quantity ${\rm S_{zt}}/c$ takes values $\sim-1$, where $c$ is 
the vacuum speed of light, whereas close to horizontal 
tracks yield values ${\rm  S_{zt}}/c \sim 0$.
The shallow ``quality bright sample'' requires an additional
condition of ${(\rm S_{zt}}/c + \cos \theta) \geq -0.4$ where $\theta$
is the reconstructed zenith angle from the LineFit. This condition excludes
events for which the one-dimensional fit suggests a significantly more
vertical downward-going geometry than the LineFit. Both signal and background are
reduced by less than $\sim$2\% by this criterion. 
Figure~\ref{fig:level2} shows the distributions of NPE (panel~(a)) and
cos~$\theta$ (panel~(b)) for experimental data, background and signal
simulations in the quality bright sample. 
The distributions of {\sc corsika-sibyll} with an iron primary
composition show a reasonable agreement with experimental data while
the total event rates are 50\% overestimated by simulation.
The zenith angle reconstruction resolution of the shallow quality
  bright sample is $\sim$1.4$^\circ$ RMS for muon bundle background and
$\geq \sim$2.5$^\circ$ for $\nu_\mu$ signal. This is because the $\nu_{\mu}$ signal experiences more
stochastic energy losses along with hadronic cascades at
its interaction vertices. 

The deep bright events (${\rm Z_{LN}}$ $\leq$ $-$300~m) are mostly events that traverse the bottom edge of IceCube or are uncontained
events that propagate or cascade below the
detector. The inclination of these events tends to be
reconstructed more horizontally than the true direction.
The agreement between the simulation and experimental distributions
improves with increasing NPE threshold values for these events. 
Events with NPE $\leq 10^4$ are discarded from the deep
quality bright sample in order to achieve a reasonable agreement between
experimental data and simulations. Since the majority of the EHE
neutrino-induced events have NPE $\geq 10^4$, the effect on the signal
efficiency by this requirement is minimal. A fraction of 96\% of
background is rejected by the cut, while 91\% of signal is retained. 
The panel~(c) in Fig.~\ref{fig:level2} shows the NPE distributions
from the deep quality bright sample. 

\subsection{Final selection}
\begin{figure}[hbt]
  \includegraphics[height=1.5in, width=1.68in]{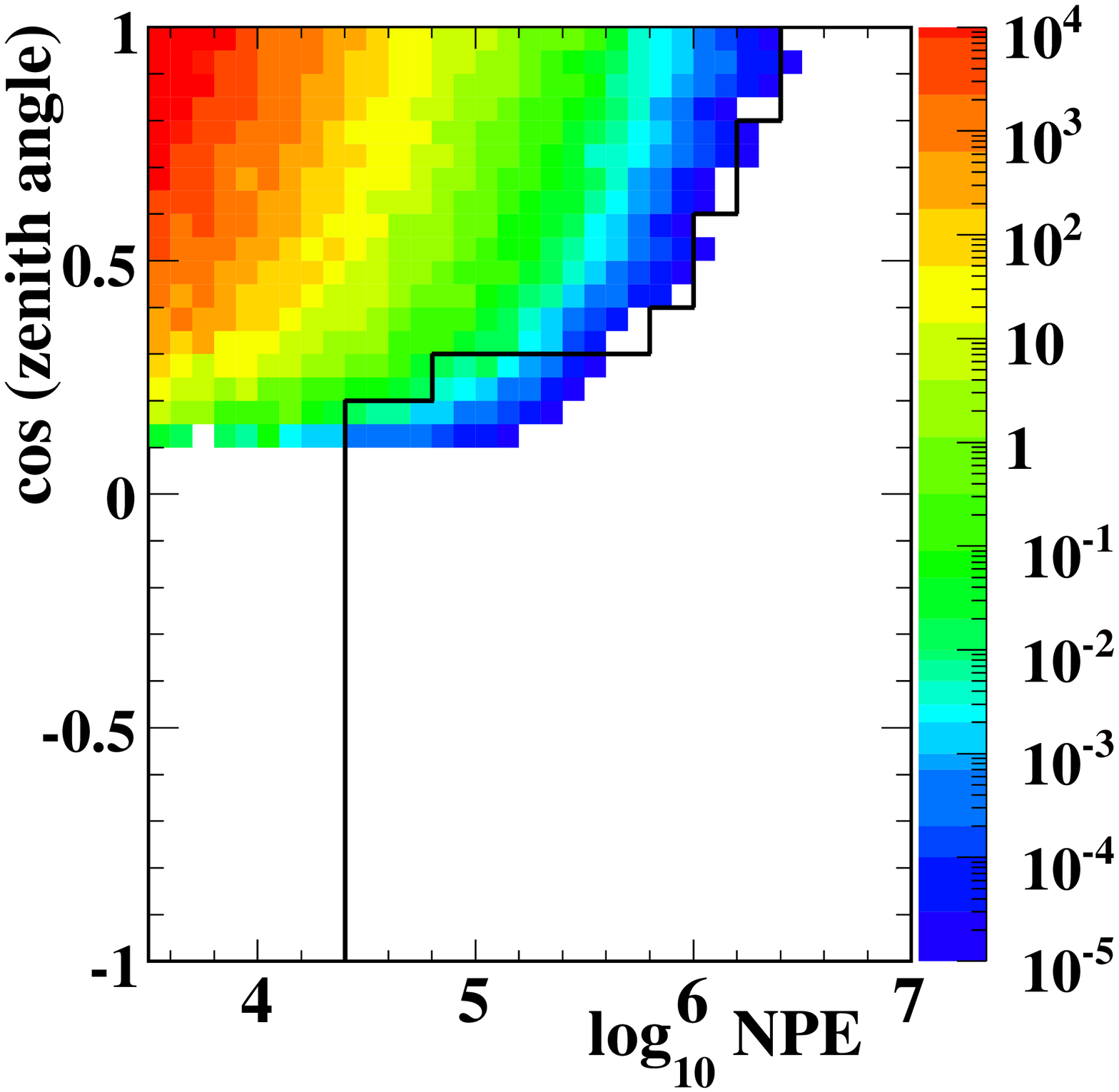}
  \includegraphics[height=1.5in, width=1.68in]{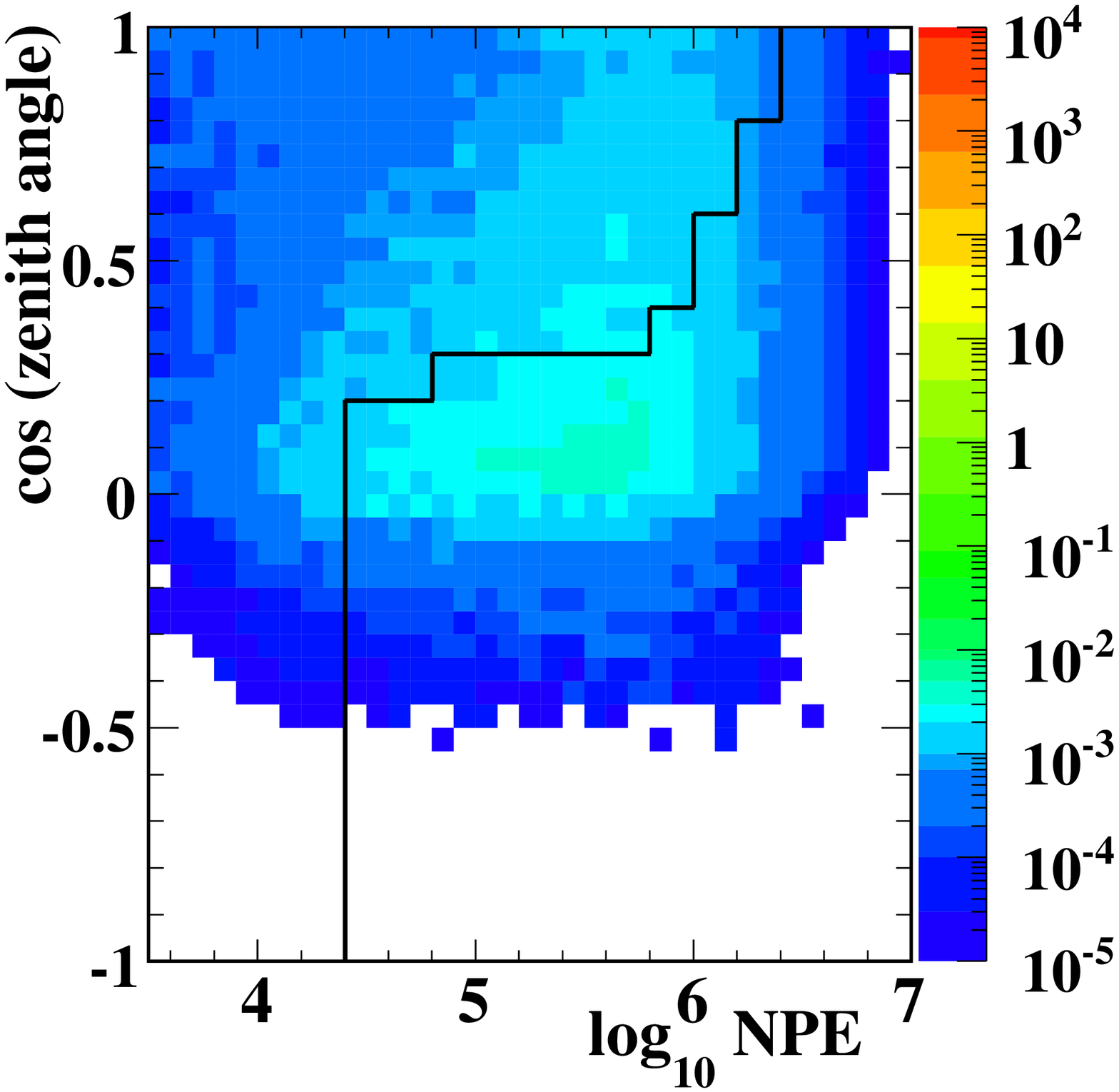}\\
  \includegraphics[height=1.5in, width=1.68in]{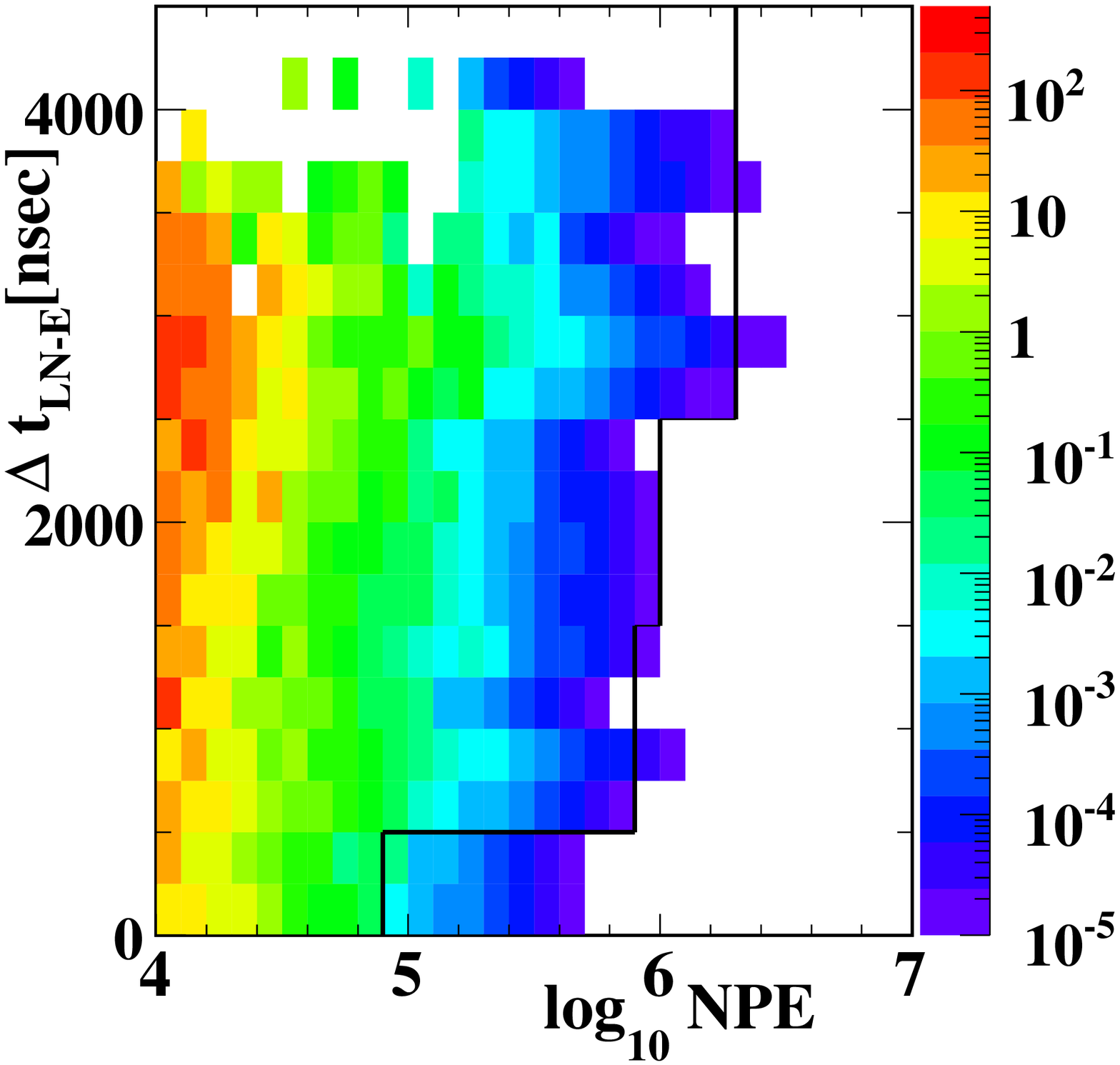}
  \includegraphics[height=1.5in, width=1.68in]{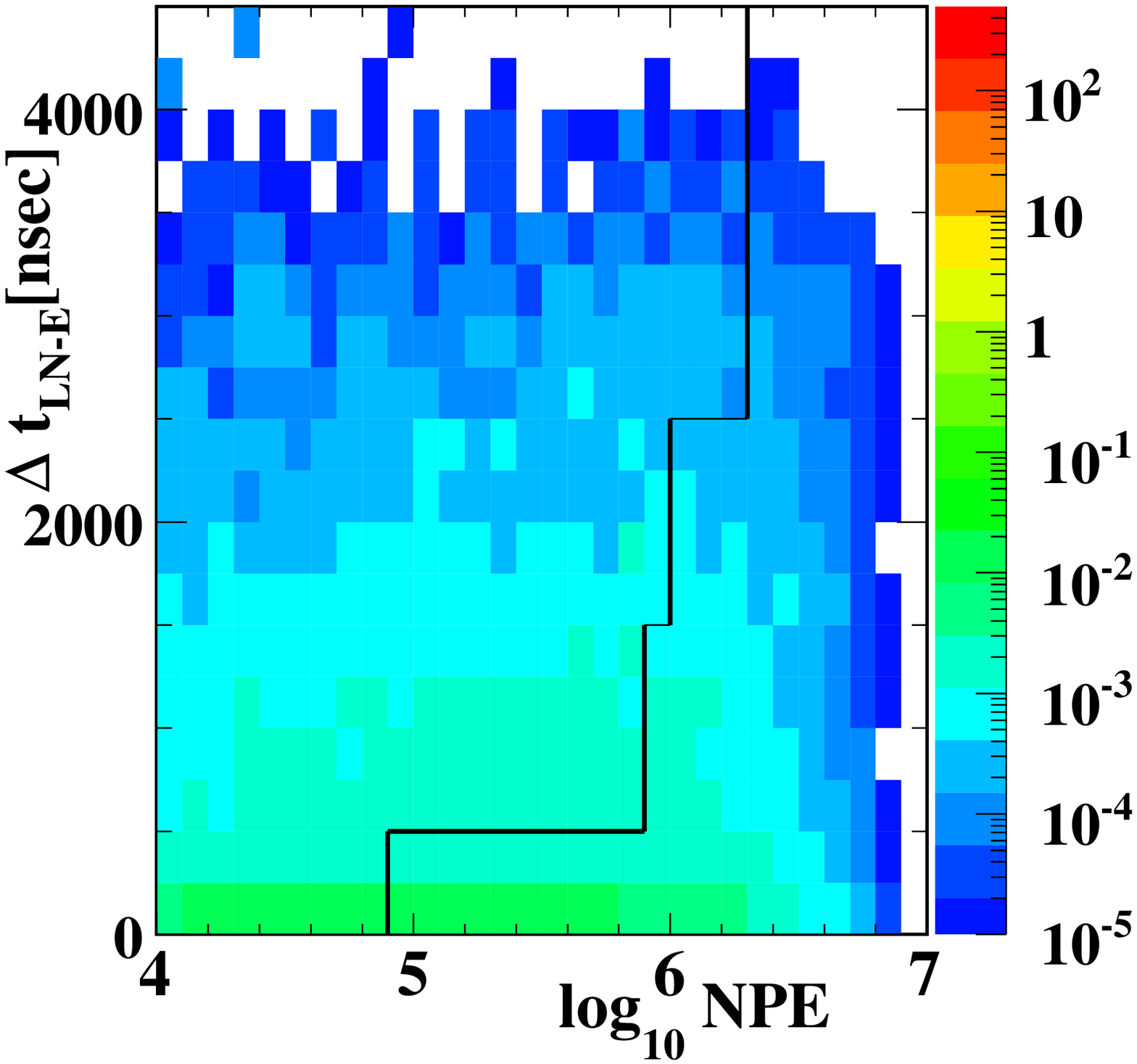}
 \caption{
Event number distributions of the shallow (upper panels) and deep
(lower panels) quality bright sample in 333.5~days are shown for the background (left
panels) and signal (right panels) simulations. The signal
distributions are from GZK~1 model~\cite{yoshida93} adding all three
flavors of neutrinos. The background distributions are from {\sc
  corsika-sibyll} with iron primaries. The series of thick lines in
each panel indicate the final sample selection criteria.
\label{fig:2D}}  
\end{figure}

The final event selection is chosen in order to minimize the model
discovery factor (MDF = $\mu_{\rm lds}/N_{\rm signal}$)~\cite{dpt} in the
region of the phase space where a better signal to background ratio
(S/B) is expected, where $\mu_{\rm lds}$ is the least number of events to
claim signal discovery at 5$\sigma$ significance and $N_{\rm signal}$ is
the number of neutrinos expected from the GZK~1~\cite{yoshida93} model
flux.
For the shallow events, high S/B is obtained in the region near
the horizontal reconstructed direction as shown in
Fig.~\ref{fig:level2} (b). For the deep events, instead of
reconstructing the inclination of events, we utilize the
time interval, ${\rm \Delta t_{LN-E}}$, between the earliest
detected photon in an event and ${\rm t_{LN}}$ to obtain the best S/B subsample.
The vertical atmospheric muon bundle events with the 
largest p.e.\ near the bottom of IceCube are often associated with a small
number of p.e.\ in the shallow detector region much earlier than 
${\rm t_{LN}}$. This contrasts to the EHE neutrino signal events. The
main contributions to a detectable EHE signal in IceCube come from neutrino-induced horizontal muons and taus~\cite{yoshida04}. 
These produce the largest p.e.\ signals shortly
after the first recorded photo-electrons. Contained cascade-like
events induced by neutrino interactions~\cite{cascade} inside the
IceCube detector volume also exhibit a similar trend. Figure~\ref{fig:level2} (d) shows the distributions
in the deep quality bright sample. The best S/B is achieved in
the bin ${\rm \Delta t_{LN-E} \sim 0}$~ns.
The high rate in the experimental data for ${\rm \Delta
  t_{LN-E} \geq 3600}$~ns is due to random noise in the DOMs and remaining coincident muons that were underestimated
by the simulations. The slightly higher rate for the data in the bin ${\rm
  \Delta t_{LN-E} \sim 0}$~ns may reflect the fact that the ice
is cleaner than what was simulated in the deep region.
Figure~\ref{fig:2D} presents the event distributions in the planes of
cos $\theta$ vs.\ NPE for the shallow events and ${\rm \Delta
  t_{LN-E}}$ vs.\ NPE for the deep events. Optimization is performed by
differentiating the NPE threshold numbers in the region cos~$\theta$ $\leq$ 0.3 or
${\rm \Delta t_{LN-E}}$ $\leq$ 0.5~$\mu$s for shallow and deep quality bright sample respectively. The NPE threshold of the other
region (cos~$\theta$ $\geq$ 0.3 or
${\rm \Delta t_{LN-E}}$ $\geq$ 0.5~$\mu$s) is conservatively determined such that the number of background
events above the threshold is less than $10^{-4}$ of the full livetime
for each bin of cos~$\theta$ with width 0.2 or 1~$\mu$s for ${\rm
  \Delta t_{LN-E}}$. This improves the detection sensitivity without
sacrificing discovery potential. The solid lines in Fig.~\ref{fig:2D} are
the final level selection criteria determined from the background
({\sc corsika-sibyll}, iron) and signal (GZK~1~\cite{yoshida93}) Monte
Carlo simulations following a blind analysis strategy. 
The minimum NPE threshold value is $2.5 \times 10^4$.
Events with NPE above the threshold value in each bin are considered to be
signal event candidates. 
No events above the threshold are found in the 10\% subset of the experimental sample.
Monte Carlo simulations indicate that a cosmic-ray primary energy of at least $\sim$2 $\times 10^{9}$~GeV is required for a muon bundle to be selected as the final sample. 
%
Table \ref{table:level2} summarizes the number of events retained in
each level of analysis.
\begin{table*}
\caption{
  Number of events passing cuts at various selection levels with
  333.5~days detector livetime. The signal rates correspond to simulations of the GZK~1
  model~\cite{yoshida93}. The errors of the on-line, off-line
  and quality bright samples are statistical only. Systematic uncertainties in the expected 
  event rates at the final selection level are given as asymmetric error 
  intervals after the statistical error.
  \label{table:level2}}
\begin{ruledtabular}
  \begin{tabular}{lrrrrrr}
    Samples
      &  \multicolumn{2}{c}{Experimental} 
      &  \multicolumn{2}{c}{Background MC ({\sc sibyll}, iron)} 
      &  \multicolumn{2}{c}{Signal MC (GZK~1)} \\
   \hline
   {\it On-line}
   &  \multicolumn{2}{c}{$3.7 \times10^7$ } 
   &  \multicolumn{2}{c}{$(3.8 \pm 0.1) \times 10^7$} 
   &  \multicolumn{2}{c}{$1.8$ $\pm 0.007$}\\
   {\it Off-line}
   &  \multicolumn{2}{c}{$3.3 \times10^5$} 
   &  \multicolumn{2}{c}{$(4.8 \pm 0.2)\times10^5$} 
   &  \multicolumn{2}{c}{$1.2 \pm 0.006$} \\
   \hline
   & Shallow & Deep & Shallow & Deep & Shallow & Deep \\ 
   \hline
   {\it Quality} & $2.9\times10^5$ & $1.9\times10^3$  & $(4.4 \pm 0.2)\times10^5$ &  $(1.7 \pm 0.2) \times10^3$  & 0.76$\pm0.005 $ & 0.43$\pm0.004 $ \\
    {\it Final} &  0  &  0  &  0.076$\pm$0.012$^{+0.051}_{-0.075}$ &  0.032$\pm$0.010$^{+0.022}_{-0.032}$ & 0.39$\pm$0.004$^{+0.054}_{-0.043}$& 0.18$\pm$0.002$^{+0.025}_{-0.020}$\\
  \end{tabular}
 \end{ruledtabular}
\end{table*}
\begin{table}
\caption{List of the statistical and systematic errors for signal
  (top) and background (bottom) simulations. The uncertainties for
  the signal are listed relative to the rate estimated for 
  GZK~1~\cite{yoshida93}. The uncertainties in the signal rates vary
  with assumed signal spectra. The uncertainties in the background rate
  are estimated with {\sc corsika-sibyll} assuming iron composition.
  \label{table:sys}}
\begin{ruledtabular}
  \begin{tabular}{lc}
    Sources
    & {Signal rate (\%)}\\ 
    \hline
        {Statistical error}
        &  {$\pm0.8$}\\
        {NPE}
        &  {+3.9 / -7.2}\\ 
        {Noise}
        &  {-1.8}\\ 
        {Neutrino cross section}
        &  {$\pm9.0$} \\
        {Photo-nuclear interaction}
        &  {+10.0} \\
        {LPM effect}
        &  {$\pm1.0$}\\ 
        \hline \multicolumn{2}{c}
    {Total:\; {$\pm0.8$(stat.) +14.0 -11.7(sys.)}}\\
    \hline
    \hline
    Sources
    & {Background rate (\%)}\\ 
    \hline
    {Statistical error}
    &  {$\pm17.0$}\\
    {NPE}
    &  {+37.1 / -46.7}\\
    {Noise}
    &  {-2.2}\\  
    {Cosmic-ray composition}
    &  {-83.9}\\ 
    {Hadronic interaction model}
    &  {+36.1} \\
    {Coincident events}
    &  {+31.2} \\
    \hline \multicolumn{2}{c}
    {Total:\; {$\pm17.0$(stat.) +60.4 -96.0(sys.)}}\\
  \end{tabular}
 \end{ruledtabular}
\end{table}
\section{The systematics}
Table~\ref{table:sys} summarizes the sources of statistical and
systematics errors in signal and background. The systematic
uncertainties are assumed to have a flat distribution and are summed
in quadrature separately for background and signal.

The dominant source of systematic uncertainty in the signal event rate is the
relationship between the measured NPE and the energy of the charged particles.
The uncertainty is estimated by calibrating the absolute sensitivity
of the DOMs in the laboratory and by calibrating the {\it in-situ} sensitivity
using light sources co-deployed with the DOMs in the ice. The estimation by the
latter method involves systematic errors in the simulation of the photon propagation in the ice.
The uncertainty associated with possible underestimation in the DOM's random
noise is estimated by adding artificial
random photo-electrons into 10\% of the simulated events.
The other uncertainties attributed to the neutrino interactions~\cite{CooperSarkar} and their daughters' interactions in the ice
are similarly estimated as in the previous analysis~\cite{ic22}.

The dominant source of systematic uncertainty in the background event rates arises from the uncertainty in  
the primary cosmic-ray composition at the relevant 
energies ($>$$10^7$~GeV) and the hadronic interaction model used in the simulation of the air showers.
The systematic uncertainty associated with the cosmic-ray composition is evaluated by 
considering two extreme cases of atmospheric muon simulations with either pure iron 
or pure proton primary compositions. 
Similarly, the uncertainty due to the hadronic interaction model is
evaluated using atmospheric muon simulations with two different high
energy hadronic interaction models: {\sc sibyll 2.1} and {\sc
  qgsjet-ii}. Systematic uncertainties associated with the NPE measurement
and the possible DOM noise rate underestimation are determined in the same manner as for signal events. 
The background contribution from possible prompt muons created in 
decays of charmed mesons is negligible.
There is also uncertainty due to statistical limitations of the simulated
coincident muon sample at NPE $\geq 10^4$. This error is estimated
by extrapolating distributions of statistically richer lower NPE
coincident simulation events to the final selection region. Possible
coincident events in the final sample are also estimated by the
temporally and geometrically separated p.e.\ signals from the main
p.e.\ cluster in each event. This coincident event check suggested that
one of the two upward-going reconstructed events in
Fig.~\ref{fig:level2}-(b) at cos~$\theta = -0.38$ was due to coincident
muons. The other upward-going event (cos~$\theta = -0.83$) was possibly
neutrino-induced, while the NPE values of both events were approximately 
4300~p.e., a factor of six less than the final threshold value.   

\section{Results}
No events in the blinded 90\% experimental data pass all the selection criteria. 
This is consistent with the expected background level of 
 $0.11\pm0.02^{+0.06}_{-0.10}$ events in a livetime of 333.5~days.
The passing rates for experimental and simulated events at each selection level are listed in Table \ref{table:level2}.
\begin{figure}
  \centering
  \includegraphics[width=3.4in]{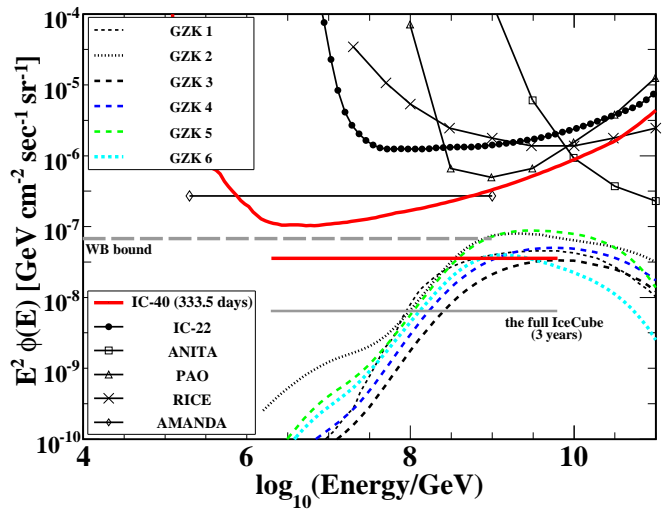}
  \caption{  All flavor neutrino flux differential limit 
  and $E^{-2}$ spectrum integrated limit from the 2008-2009 IceCube EHE 
  analysis (red solid lines). The systematic errors
  are included. Various model predictions (assuming primary protons) are shown for comparison:
  GZK~1 ((m, ${\rm Z_{max}}$) = (4,4))~\cite{yoshida93}, 
  GZK~2~\cite{kalashev02}, 
  GZK~3 ($\Omega_{\Lambda}$ = 0.0), 
  GZK~4 ($\Omega_{\Lambda}$ = 0.7)~\cite{ESS}, 
  GZK~5 (maximal) and GZK~6 (the best fit, incorporating the Fermi-LAT bound)~\cite{ahlers}. 
  The gray dashed horizontal line indicates the Waxman-Bahcall flux bound with 
  cosmological evolution~\cite{wb}.
  Model fluxes are summed over all neutrino flavors, assuming standard neutrino oscillations.
  The model independent differential upper limits by other experiments are 
  also shown 
  for Auger (PAO)~\cite{auger},
  RICE~\cite{rice}, ANITA~\cite{anita}, and the previous IceCube result (IC22)~\cite{ic22}.
  Limits from other experiments are converted to the all flavor limit assuming standard neutrino oscillation and a 90\% quasi-differential limit when  necessary.
  The integral flux limit on a pure $E^{-2}$ spectrum is shown for AMANDA-II~\cite{amanda}.
  For reference, the estimated integrated limit for three years of observation with the full IceCube detector with the same analysis strategy is denoted as a gray solid line.
  \label{fig:sensitivity}
  }
\end{figure}
\begin{figure}[hbt]
  \includegraphics[height=2in, width=1.68in]{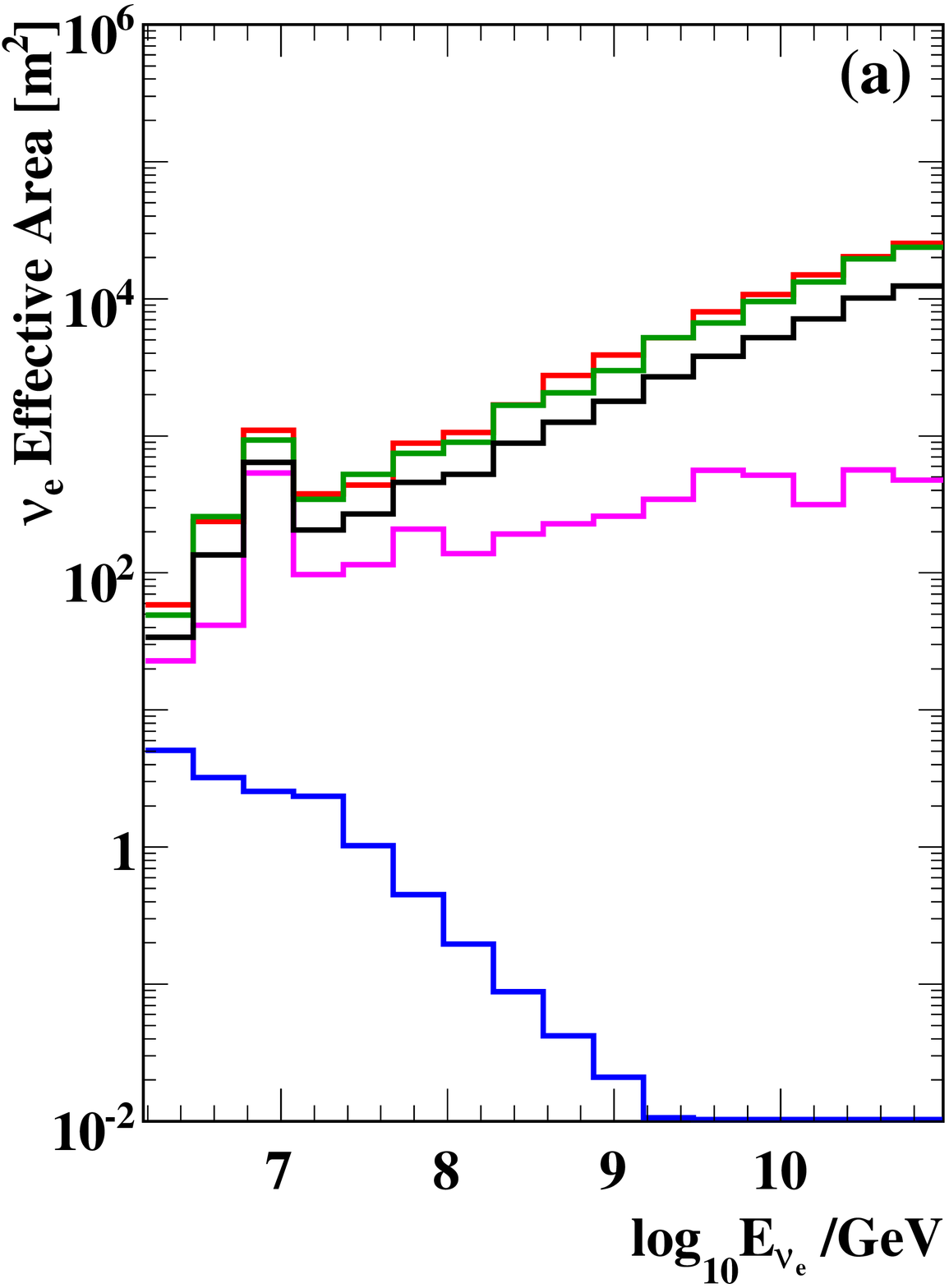}
  \includegraphics[height=2in, width=1.68in]{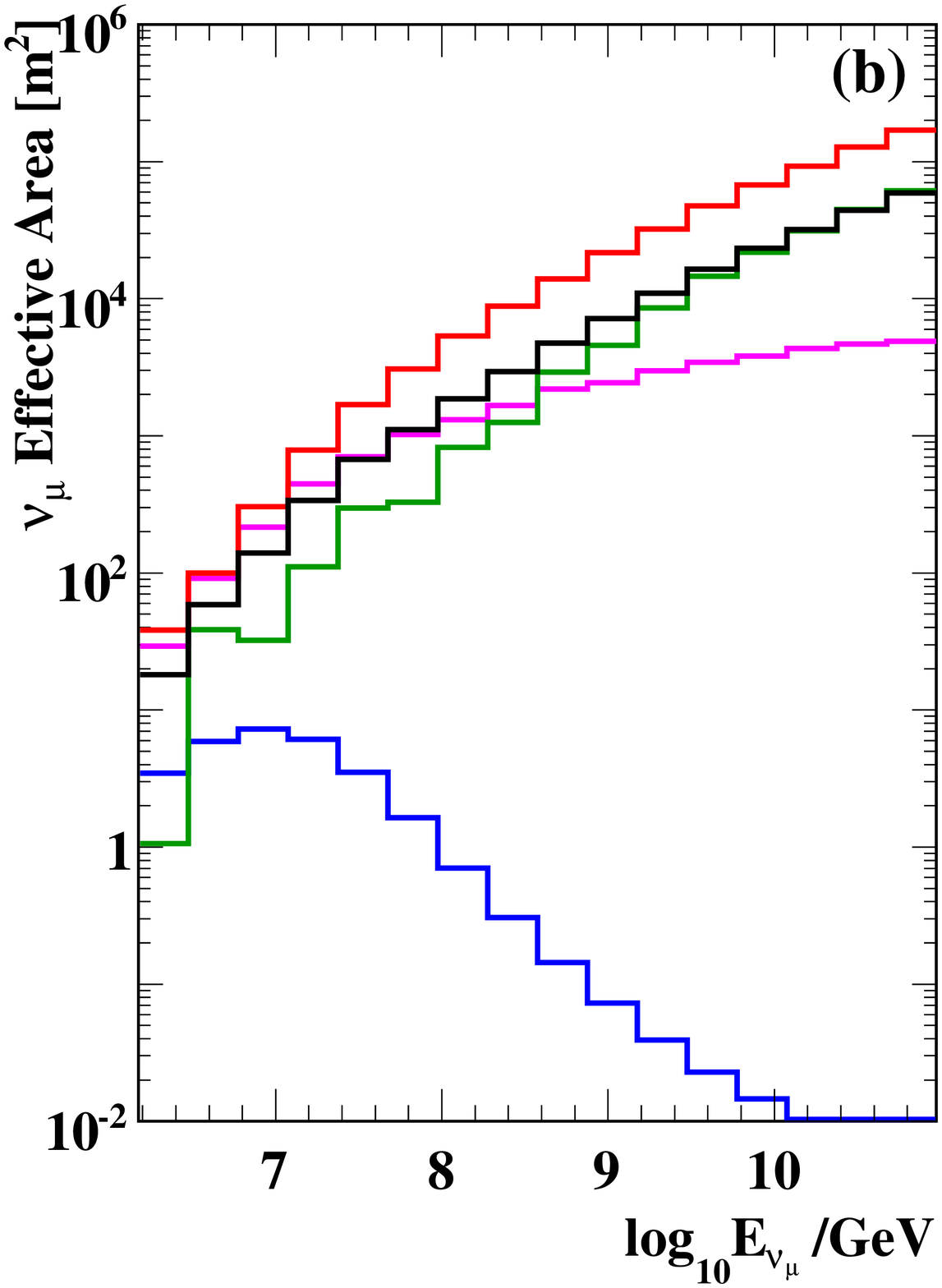}\\
  \includegraphics[height=2in, width=1.68in]{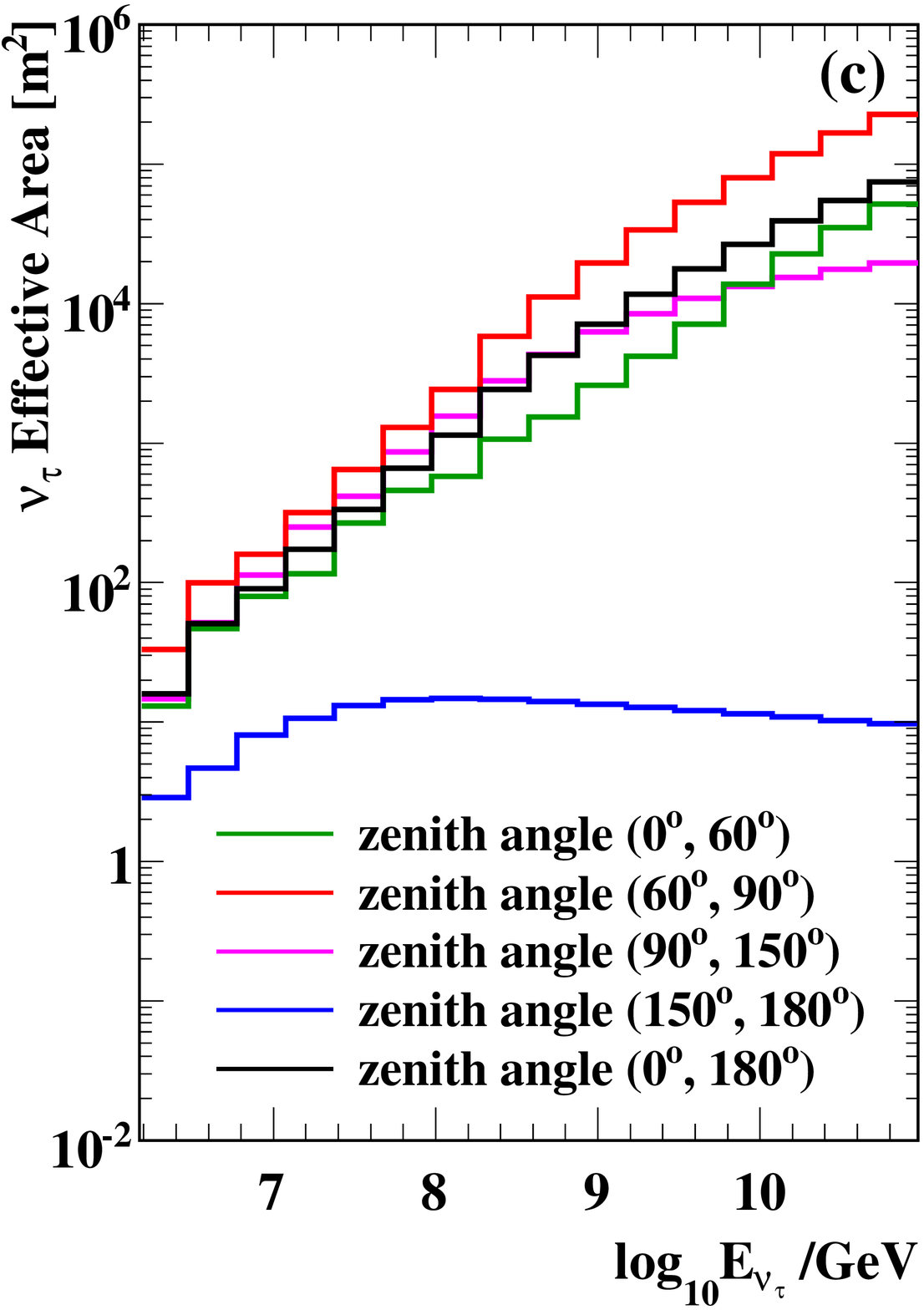}
 \includegraphics[height=2in, width=1.68in]{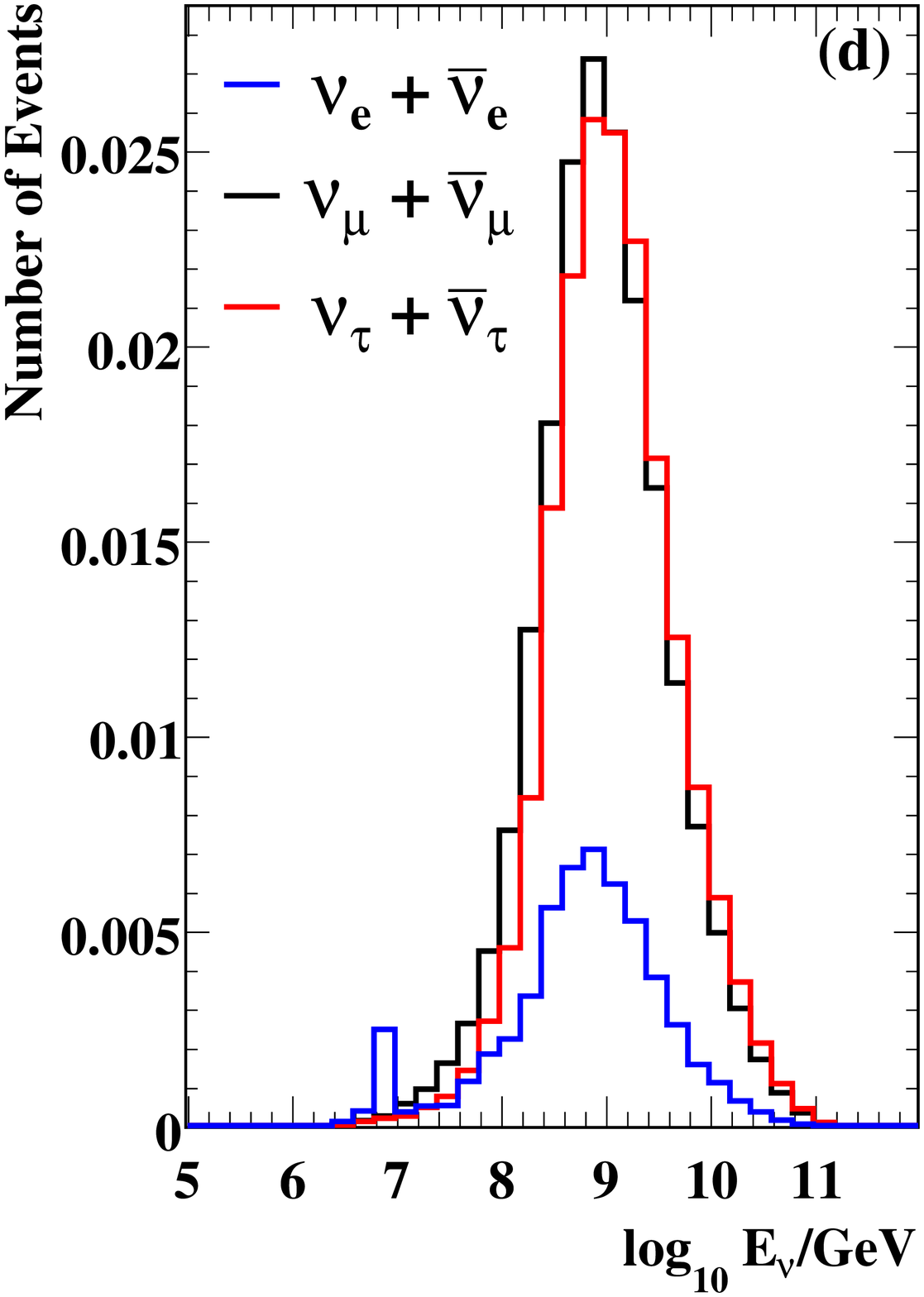}  
  \caption{Solid angle averaged neutrino effective area for four declination 
    bands as well as that of the full solid angle average
    for (a)~$\nu_{\it{e}}+\bar{\nu}_{\it{e}}$,
    (b)~$\nu_{\mu}+\bar{\nu}_{\mu}$, and (c)~$\nu_{\tau}+\bar{\nu}_{\tau}$,
    assuming equal flux of neutrinos and anti-neutrinos.
    The lower right plot shows the final level signal event distributions 
    for 333.5~days with the GZK~6 model spectra~\cite{ahlers}
    for each neutrino flavor.
    \label{fig:effArea}}
\end{figure}
\begin{table}
  \caption{Expected numbers of events in 333.5 days from several cosmogenic 
    neutrino models and top-down models.
    The confidence 
    interval for exclusion by this observations 
    is also listed where appropriate.
    The cosmogenic neutrino models (GZK~1-6) assume the cosmic-ray primaries 
    to be protons and different spectral indices/cutoff energies at sources as well as
    different cosmological evolution parameters and extension in redshift for the sources.
    Representative models with moderate (GZK~3, 4, 6), moderately strong 
    (GZK~1) and strong (GZK~2, 5) source evolution parameters are listed here.
    \label{table:final}}
  \begin{ruledtabular}
    \begin{tabular}{lcc}
      Models
      & {Event rate}
      & C.L. \%\\ 
      \hline
          {GZK~1~\cite{yoshida93}}
          & 0.57 
          & $\cdots$ \\ 
          {GZK~2~\cite{kalashev02}}
          & 0.91 
          & 53.4 \\ 
          {GZK~3 ($\Omega_{\Lambda}$ = 0.0)~\cite{ESS}}
          & 0.29
          & $\cdots$ \\ 
          {GZK~4 ($\Omega_{\Lambda}$ = 0.7)~\cite{ESS}}
          & 0.47
          & $\cdots$ \\ 
          {GZK~5 (maximal)~\cite{ahlers}}
          & 0.89
          & 52.8 \\ 
          {GZK~6 (the best fit)~\cite{ahlers}}
          &0.43
          &  $\cdots$ \\ 
          {Top-down 1 (SUSY)~\cite{sigl}}
          & 1.0
          & 55.7 \\ 
          {Top-down 2 (no-SUSY)~\cite{sigl}}
          &  5.7
          &  99.6 \\ 
          {Z-burst~\cite{yoshida98}}
          &  1.2
          &  66.4 \\ 
          {WB bound (with evolution)~\cite{wb}}
          &  4.5
          &  $\cdots$ \\ 
          {WB bound (without evolution)~\cite{wb}}
          &  1.0
          &  $\cdots$ \\ 
    \end{tabular}
  \end{ruledtabular}
\end{table}

The quasi-differential model-independent 90\% CL limit on neutrino
fluxes~\cite{anchordoqui} normalized by energy decade is shown in
Fig.~\ref{fig:sensitivity} assuming full mixing in the standard neutrino flavor
oscillation scenario.
In the limit calculation, the energy decade averaged effective area is used and the contribution from the Glashow resonance~\cite{glashow} is neglected.
Incorporating the statistical and systematic uncertainties, the background is expected to be found with a uniform
prior probability between 0 and 0.19. This uncertainties are included in the final limit using a
method outlined in \cite{pole++}.
This estimation together
with the null result in the experimental sample gives the
Feldman-Cousins 90\% CL event upper limit~\cite{feldman98} of 2.35 events. 
For cosmic neutrinos with an $E^{-2}$ energy spectrum, this implies an integral flux limit of $E^2 \phi \leq 3.6 \times 10^{-8}$~${\rm GeV cm^{-2} sec^{-1}sr^{-1}}$ with the central 90\% of the $E^{-2}$ signal found in the energy interval $2.0 \times 10^{6}$ $-$ $6.3 \times 10^{9}$~GeV\null.
This result is the first constraint of neutrino fluxes below the Waxman-Bahcall flux bound~\cite{wb} in this energy region.

\section{Discussions and Summary}
%
We analyzed the 2008-09 data sample collected by the 40-string IceCube detector to search for extremely-high energy neutrinos with energies exceeding $10^6$~GeV\null. 
The differential and integral limits obtained are significantly improved relative to our previous result~\cite{ic22}. 
This is due to both the increased instrumented volume and 
improvements of the Monte Carlo simulations. The improved agreement 
between experimental and simulated data allowed a loosening of the NPE threshold in 
the data selection, thereby lowering the energy threshold of the analysis 
and improving the selection efficiency for high energy signal events that 
occurred outside the instrumented volume.
%
This can also be seen in the corresponding neutrino effective area at 
the final selection shown in Fig.~\ref{fig:effArea}. Compared to the 
previous search~\cite{ic22}, the effective area is a factor of 6 and 
3.3 increased at $3 \times 10^7$~GeV and $10^9$~GeV, respectively.
The full solid angle averaged 3 flavor (assuming
$\nu_e$:$\nu_\mu$:$\nu_\tau$=1:1:1) neutrino effective area reaches 300~m$^2$ at $10^8$~GeV and 2100~m$^2$ at $10^9$~GeV\null.
The 90\% CL differential limit at $10^9$~GeV is a factor of
$\sim$4 higher than the fluxes predicted by the models GZK~2 and 5, and a factor of $\sim$8
higher than the flux predicted by the models GZK~1, 4, 6, all of  
which assume primary protons. This suggests that the IceCube EHE
neutrino search will reach these flux levels in the near future
since the event rate is roughly proportional to the fiducial volume
(see Fig.~\ref{fig:sensitivity}), and the current analysis used only the half-instrumented IceCube detector configuration. Further improvements in sensitivity would enable  
IceCube to act as a a probe of the primary cosmic-ray composition at  
GZK energies~\cite{Anchordoqui:2007fi}.

Figure~\ref{fig:effArea} indicates that a large part of the EHE neutrino
signal are expected from the zenith angle region between $60^{\rm o}$
and $90^{\rm o}$. Upward-going EHE neutrinos are absorbed in
the Earth. The propagation length of secondary muons and taus is greater
than the distance between the surface and the IceCube fiducial volume. Thus,
the inclined particles that reach the IceCube detector are created in the Earth.
For $\nu_e$, the event signatures are produced nearly
at the neutrino interaction points and the current analysis is sensitive to all downward-going geometries.
The peaked features in Fig.~4 (a) and (d) at $E_{\nu_e} \sim$ 6.3~PeV is due to the Glashow resonance~\cite{glashow}.
Expected signal energy distributions of GZK~6 at the final selection level
are shown in the lower right panel in Fig.~\ref{fig:effArea}. The
peak energy of the expected signal after all selection criteria is at $\sim$7.0 $\times
10^8$~GeV\null. Significant contributions from all neutrino flavors are
observed. In the GZK~6 model, 13\% of the signal are from $\nu_e$, 45\% are
from $\nu_\mu$ and 42\% are from $\nu_{\tau}$.
Through-going tracks (muons and taus) constitute 60\% of the signal
rate and the rest are neutrino interactions that create cascade-like events
near and inside the detector volume.
Table~\ref{table:final} gives the event rates for several
model fluxes of cosmogenic neutrinos, top-down scenarios, and a pure $E^{-2}$ power-law
neutrino spectrum normalized to the Waxman-Bahcall flux bounds for reference.
We expect 0.3 to 0.9 cosmogenic neutrino events in 333.5~days, 
assuming moderate to strong cosmological source evolution models.
The half-instrumented IceCube detector is already capable of constraining those models with relatively
high neutrino fluxes. 
The IceCube sensitivity to cosmological EHE neutrinos continues to grow.

\begin{acknowledgments}
We acknowledge the support from the following agencies:
U.S. National Science Foundation-Office of Polar Programs,
U.S. National Science Foundation-Physics Division,
University of Wisconsin Almni Research Foundation,
the Grid Laboratory Of Wisconsin (GLOW) grid infrastructure at the University of Wisconsin - Madison, the Open Science Grid (OSG) grid infrastructure;
U.S. Department of Energy, and National Energy Research Scientific Computing Center,
the Louisiana Optical Network Initiative (LONI) grid computing resources;
National Science and Engineering Research Council of Canada;
Swedish Research Council,
Swedish Polar Research Secretariat,
Swedish National Infrastructure for Computing (SNIC),
and Knut and Alice Wallenberg Foundation, Sweden;
German Ministry for Education and Research (BMBF),
Deutsche Forschungsgemeinschaft (DFG),
Research Department of Plasmas with Complex Interactions (Bochum), Germany;
Fund for Scientific Research (FNRS-FWO),
FWO Odysseus programme,
Flanders Institute to encourage scientific and technological research in industry (IWT),
Belgian Federal Science Policy Office (Belspo);
University of Oxford, United Kingdom;
Marsden Fund, New Zealand;
Japan Society for Promotion of Science (JSPS);
the Swiss National Science Foundation (SNSF), Switzerland;
A.~Gro{\ss} acknowledges support by the EU Marie Curie OIF Program;
J.~P.~Rodrigues acknowledges support by the Capes Foundation, Ministry of Education of Brazil.
\end{acknowledgments}

\newpage
{\bf \large 
Erratum: Constraints on the Extremely-high Energy Cosmic Neutrino Flux with the IceCube 2008-2009 Data}\\

\begin{figure}[h]
  \includegraphics[height=2in, width=1.68in]{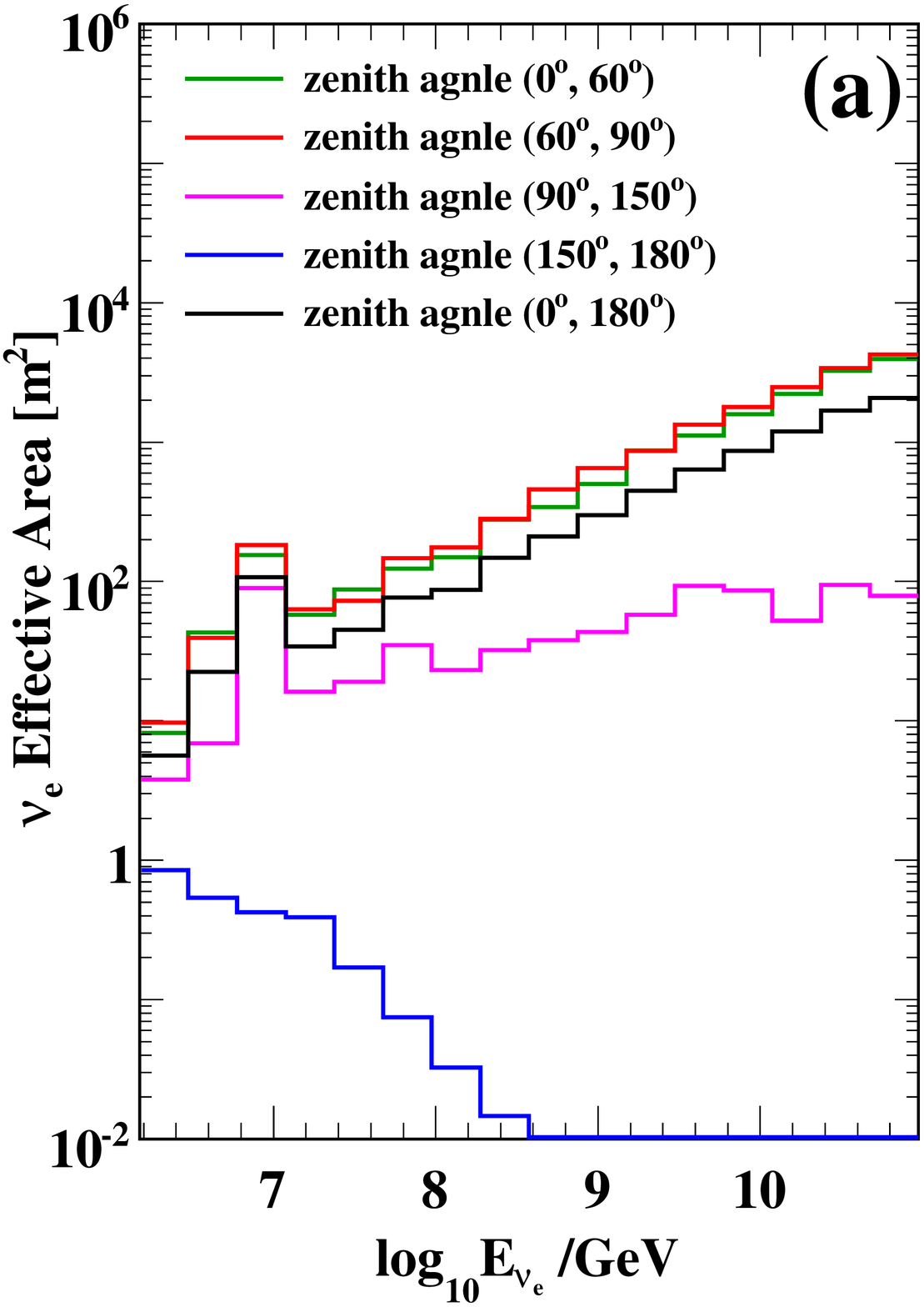}
  \includegraphics[height=2in, width=1.68in]{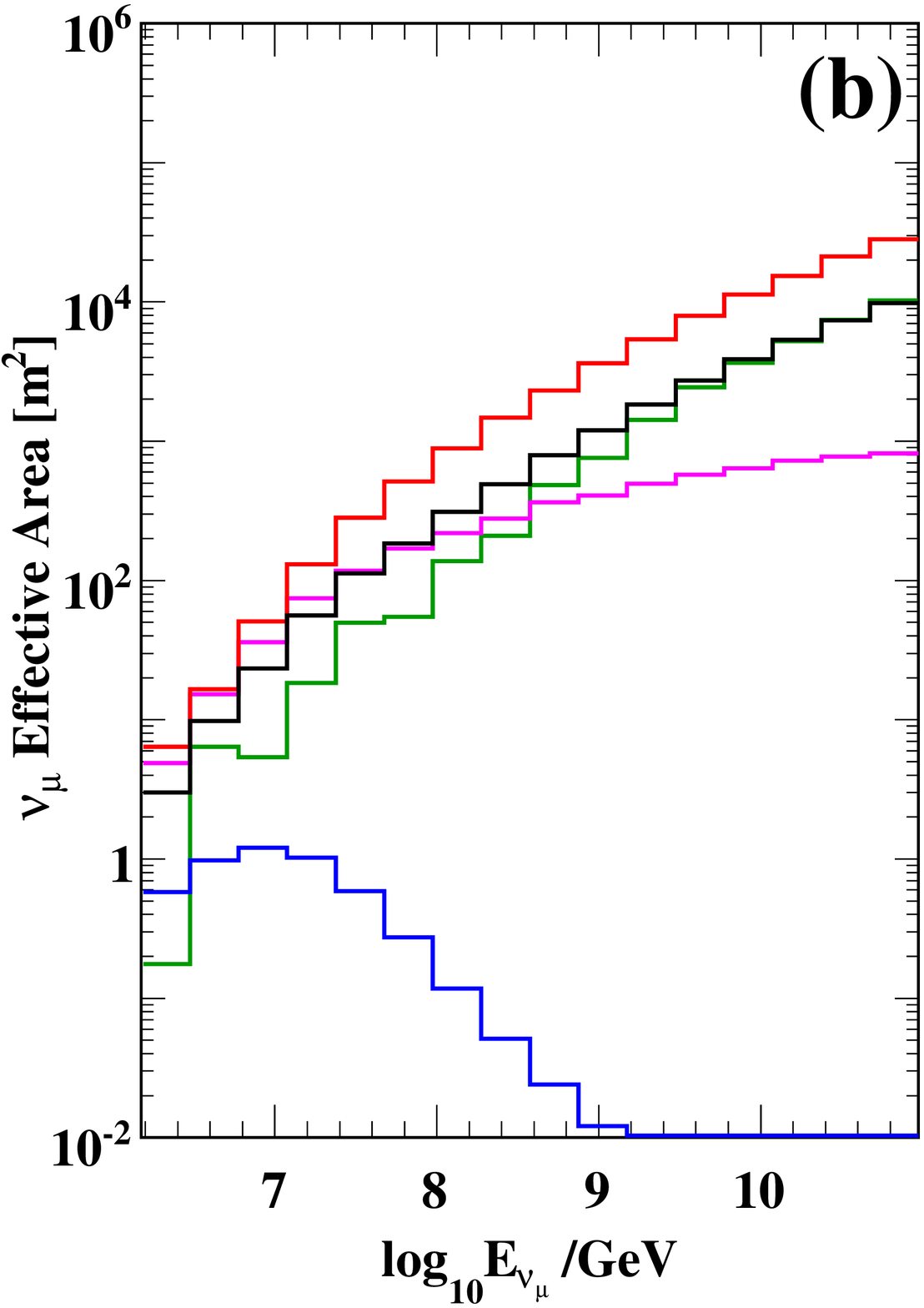}\\
  \includegraphics[height=2in, width=1.68in]{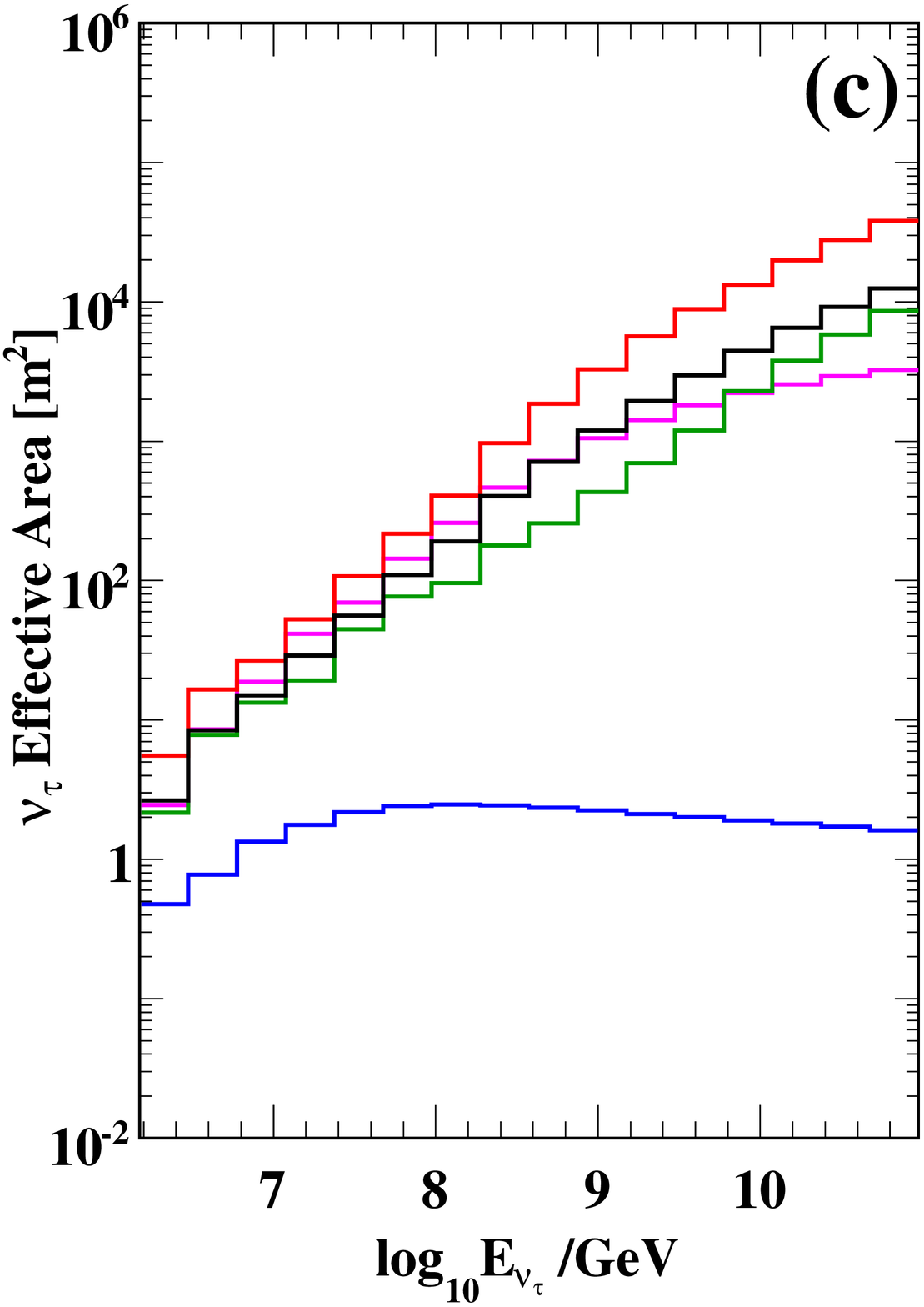}
  \includegraphics[height=2in, width=1.68in]{Fig4_null.eps}
  \caption{Solid angle averaged neutrino effective area for four declination 
    bands as well as that of the full solid angle
    for (a)~$\nu_{\it{e}}+\bar{\nu}_{\it{e}}$,
    (b)~$\nu_{\mu}+\bar{\nu}_{\mu}$, and (c)~$\nu_{\tau}+\bar{\nu}_{\tau}$,
    assuming equal flux of neutrinos and anti-neutrinos.
    \label{fig:area}}
\end{figure}

An error has been found in the presentation of the neutrino effective areas in 
Figure 4 (a)--(c) of the original paper, which led to an overestimation of the values by a factor of six.
Corrected neutrino effective areas are shown in Fig.~\ref{fig:area}.
All other results reported in the paper, including the upper limit, were not affected and hence remain unchanged.
\end{document}